\documentclass[12pt]{iopart}
\usepackage{iopams}
\usepackage{graphicx}
\usepackage{amssymb}
\usepackage{multirow}
\usepackage{subfig}
\usepackage{color}
\usepackage{epstopdf}
\usepackage{leftidx}
\usepackage{arydshln}
\usepackage[font=small,labelfont=bf]{caption}
\usepackage[nospace]{cite}

\begin{document}
	
\title[Target-based Optimization of Advanced Gravitational-Wave Detector Networks]{Target-based Optimization of Advanced Gravitational-Wave Detector Network Operations}

\author{\'A. Sz\"olgy\'en$^{1,2}$, G. D\'alya$^{1,2}$, L. Gond\'an$^{1,2}$, P. Raffai$^{1,2}$}
\address{$^1$ E\"otv\"os University, Institute of Physics, 1117 Budapest, Hungary\\}
\address{$^2$ MTA-ELTE EIRSA ``Lend\"ulet'' Astrophysics Research Group, 1117 Budapest, Hungary\\}
\ead{szolgyen@caesar.elte.hu}

\begin{abstract}
We introduce two novel time-dependent figures of merit for both online and offline optimizations of advanced gravitational-wave (GW) detector network operations with respect to (i) detecting continuous signals from known source locations and (ii) detecting GWs of neutron star binary coalescences from known local galaxies, which thereby have the highest potential for electromagnetic counterpart detection. For each of these scientific goals, we characterize an $N$-detector network, and all its $(N-1)$-detector subnetworks, to identify subnetworks and individual detectors (key contributors) that contribute the most to achieving the scientific goal. Our results show that aLIGO-Hanford is expected to be the key contributor in 2017 to the goal of detecting GWs from the Crab pulsar within the network of LIGO and Virgo detectors. For the same time period and for the same network, both LIGO detectors are key contributors to the goal of detecting GWs from the Vela pulsar, as well as to detecting signals from 10 high interest pulsars. Key contributors to detecting continuous GWs from the Galactic Center can only be identified for finite time intervals within each sidereal day with either the 3-detector network of the LIGO and Virgo detectors in 2017, or the 4-detector network of the LIGO, Virgo, and KAGRA detectors in 2019-2020. Characterization of the LIGO-Virgo detectors with respect to goal (ii) identified the two LIGO detectors as key contributors. Additionally, for all analyses, we identify time periods within a day when lock losses or scheduled service operations could result with the least amount of signal-to-noise or transient detection probability loss for a detector network.

\end{abstract}


\submitto{\CQG}

\maketitle

\section{Introduction}
\label{sec1}

Second generation (advanced) gravitational-wave (GW) detectors are state-of-the-art, ground-based, L-shaped interferometers with kilometer-scale arms. Such detectors include Advanced LIGO (aLIGO) \cite{Abbott2015}, whose first observing run started in 2015, and Advanced Virgo (AdV) \cite{Acernese2015} which is expected to start its first observing run in 2017 \cite{ADEOverview}. Two additional GW detectors are planned to join the network of aLIGO and AdV: (i) the Japanese KAGRA is under construction with baseline operations beginning in 2018 \cite{Somiya2012}, while (ii) the proposed LIGO-India is expected to become operational in 2022 \cite{Iyer2011,Abbott2016}. Studies have already suggested site locations and orientations of arms for LIGO-India based on scientific figures of merit \cite{Raffai2013}, however, these parameters have still not been finalized yet. 

Although advanced GW detectors are capable of detecting GWs from the whole sky, their directional sensitivity is anisotropic \cite{Schutz2011}. As the simultaneous operation of multiple GW detectors allows coherent detection of a GW from the same astrophysical source, the combined directional sensitivity of a network of detectors can also be defined \cite{Schutz2011}. The co-rotation of the directional sensitivity pattern of a network with Earth results with a periodic change in the signal-to-noise ratio (SNR) detectable by the network from a given continuous source, as well as in the transient detection rate of a network from given hosts. By knowing the location of a specific source or host in the sky, we can quantify the directional sensitivity of a network towards the source or host at different times of the day, and by choosing a specific source and waveform model, we can translate this to SNR or transient detection rate observable by the network with the corresponding sensitivities. As we will show in section \ref{sec2} and \ref{sec3}, this characterization of a GW detector network can be generalized to multiple sources or hosts as well.

Design operation timelines, design strain and directional sensitivities of individual GW detectors in a network imply a set of \textit{scientific goals} that, according to reasonable expectations derived from source models, can potentially be achieved with the network. In case of advanced GW detector networks, such scientific goals include detecting continuous GWs from high interest pulsars \cite{WhitePaper2014,Aasi2014,Ming2016} or from the Galactic Center \cite{WhitePaper2014,Aasi2013b}, or achieving joint detections of coalescing binaries with GW detectors and electromagnetic (EM) telescopes \cite{WhitePaper2014,Somiya2012,Hanna2014}. For searches for continuous GWs directed to selected pulsars or to the Galactic Center, the SNR collected within an observing time $\Delta t$ satisfying $T_\mathrm{GW} \ll \Delta t\ll T_\mathrm{Earth}$ ($T_\mathrm{GW}$ being the period of the continuous GW and $T_\mathrm{Earth}$ being Earth's rotational period) can be used as a base for a time-dependent figure of merit (FoM) characterizing a GW detector network. Alternatively, the expected number of GW detections within a $\Delta t\ll T_\mathrm{Earth}$ of neutron star binary coalescences which have the potential for triggering EM counterpart detections, can be used as a FoM 
characterizing a GW detector network in terms of its capability to ensure joint GW+EM detections of compact binary coalescences. By introducing these FoMs, we can compare different GW detector networks with respect to chosen scientific goals, independently from $\Delta t$. Such comparisons can form a basis for strategic decisions on operations e.g.\ how to schedule service operations, detector commissions, engineering runs and down times of different detectors of the network, or how to choose an improvement plan for specific detectors and for the network as a whole. Alternatively, in case of a fixed operation and improvement schedule, these comparisons can support identifying the set of scientific goals the specific schedule is optimal for. Furthermore, by comparing all the different $(N-1)$-detector subnetworks of an $N$-detector network, we can identify individual detectors which spend most of their design operation time in subnetworks that contribute the most to achieving a given scientific goal (from now on, we will refer to such a detector as the \textit{key contributor} to a scientific goal defined for the network). At any given time during a network operation, a prompt comparison between the actual and design status of the network can tell how efficient the network is at a given time in terms of a specific scientific goal, and thus it can serve as an online diagnostic tool characterizing the network. Finally, as there are proposed motivations and techniques for prompt adjustments of detector sensitivities during future operations (e.g. switching detectors between broad-band (detuned) and narrow-band (tuned) operating modes; see \cite{LCGT2009,InstWhitePaper2015} and references therein), optimizing the timing of such sensitivity adjustments with respect to specific scientific goals will be of great importance. The FoMs and characterization techniques we demonstrate in this paper are well suited for carrying out such optimizations as well.

In the analyses presented in this paper, we use the proposed operating timeline of advanced GW detector networks outlined in \cite{Abbott2016}, and select scientific goals to be achieved with these networks based on the aLIGO 2015-2016 white paper \cite{WhitePaper2014} and the 2013 KAGRA data analysis white paper \cite{WhitePaper2013KAGRA}. For each network of $N$ detectors and for their $(N-1)$-detector subnetworks, we calculate how the FoMs corresponding to different scientific goals change with time within a day. We also identify subnetworks that contribute the most to the network FoM at different time intervals. Finally, using these results, we identify the key contributors within the networks to the selected scientific goals, and suggest optimal operation strategies. Choosing a technical improvement plan for specific detectors based on our results, as well as taking into account plans for prompt adjustments of detector sensitivities in the future, are beyond the scope of this paper.

The paper is organized as follows. In section \ref{sec2}, we introduce the FoM we use to characterize GW detector networks in terms of detecting GWs from various continuous sources. In section \ref{sec3}, we describe the FoM associated with detecting GWs of neutron star binary coalescences from known hosts which has the potential for triggering targeted electromagnetic follow-up observations. Additionally, in both section \ref{sec2} and \ref{sec3}, we identify key contributors and associate them to time intervals when they contribute the most to achieving a given scientific goal. Using these information, we suggest optimal strategies for scheduling short-term services. In section \ref{sec4}, we summarize and discuss our results.

\section{Characterizing networks regarding targeted searches of continuous GWs}
\label{sec2} 

In this section we introduce the FoM we use to characterize networks with respect to detecting continuous GWs. We apply this FoM on various networks consisting of aLIGO, AdV, and KAGRA detectors, and characterize their efficiencies in detecting  continuous GWs from the Crab and Vela pulsars (section \ref{sec21}), from 10 high interest pulsars (section \ref{sec22}), and from the Galactic Center (section \ref{sec23}). 

We chose our FoM to be the ratio of SNR squares collected by a network and various subnetworks within an integration time of $\Delta t$ (satisfying $T_\mathrm{GW} \ll \Delta t\ll T_\mathrm{Earth}$) around detection time $t$. Note that the ratio of SNR squares as an FoM has the advantage of being linearly additive. We write this ratio of SNRs as $\rho_{\mathcal{N}\mathrm{sub}}^2(t) / \rho_{\mathcal{N}}^2 (t)$, where $\rho_{\mathcal{N}} (t)$ denotes the SNR collected by an $N$-detector network, and $\rho_{\mathcal{N}\mathrm{sub}}(t)$ denotes the SNR collected by a chosen ($N-1$)-detector subnetwork, both calculated for a specific source or sources of continuous GWs within a $\Delta t$ time window centered at any time $t$ of a day. Indices $\mathcal{N}$ and $\mathcal{N}_\mathrm{sub}$ denote the set of detectors in the corresponding network and subnetwork, respectively. In our analyses, we use the definition of the network SNR given in \cite{Schutz2011}, however, we also take into consideration that both $\rho_\mathcal{N}(t)$ and $\rho_{\mathcal{N}\mathrm{sub}}(t)$ are functions of time that change with Earth's rotation. The definition of $\rho_\mathcal{N}(t)$, as well as of $\rho_{\mathcal{N}\mathrm{sub}}(t)$, is given by the following expression:
\begin{equation}
\label{eq1} 
\rho_{\mathcal{N}}(t) = \sqrt{4 \sum^{N}_{k=1} \int\limits_0^{+\infty}{\frac{| \widetilde{h}_{k} (f,t) |^2}{S_{k}(f)} \mathrm{d}f}  } \, ,
\end{equation}
where ${S_{k}(f)}$ is the one-sided power spectral density (PSD) of the $k$-th detector's noise, and $\widetilde{h}_{k} (f,t)$ is the one-sided frequency spectrum of the continuous GW integrated with the $k$-th detector within a time window of $\Delta t$ around $t$. $\widetilde{h}_{k} (f,t)$ is the linear combination of $(+)$ and $(\times)$ polarizations of the GW denoted by $\widetilde{h}_+(f)$ and $\widetilde{h}_{\times}(f)$, respectively:
\begin{equation}
\label{eq2}
\widetilde{h}_{k}(f,t) = F_{+,k}(t,\psi) \widetilde{h}_+(f) + F_{\times,k}(t,\psi) \widetilde{h}_{\times}(f) \, .
\end{equation}
Here $F_{+,k}(t,\psi)$ and $F_{\times,k}(t,\psi)$ are the so-called \textit{antenna pattern functions} for the $k$-th detector which are traditionally expressed as functions of the sky coordinates of the source, and of the reference angle for the GW's polarization $(\psi)$, see formulas in \cite{Schutz2011}. Since the sky coordinates of the source in an Earth-centered frame can be calculated at any time of the day, we can indicate the time-dependence of the antenna pattern functions directly, instead of indicating their dependence on spherical coordinates of the source position in the sky. $\psi$ is practically unknown, because the GW's polarization itself is typically unknown, and thus, we average the SNR (denoted as $\overline{\rho}_{\mathcal{N}}$) over the GW's polarization to obtain the following expression which is independent from $\psi$ (see detailed steps in \cite{Schutz2011}):
\begin{equation}
\label{eq3}
\overline{\rho}_{\mathcal{N}}^2(t) = 2 \sum^{N}_{k=1} \left[ F_{+,k}^{2} (t) + F_{\times,k}^{2}(t) \right]  \int\limits_{0}^{+\infty}{\frac{| \widetilde{h} (f) |^2}{S_{k}(f)} \mathrm{d}f} \, ,
\end{equation}
where the sum of $F_{+,k}^2(t)$ and $F_{\times,k}^2(t)$ is the network's combined antenna power pattern at time $t$ towards the sky direction of the source which is independent from $\psi$ \cite{Schutz2011}. On the right hand side of Eq.~(\ref{eq3}), we assume that the integral can be solved, and it has a nonzero, finite value in case of a plausible continuous waveform that is expressed as $| \widetilde{h} (f) |^2 = |\widetilde{h}_+(f)|^2 + |\widetilde{h}_{\times}(f)|^2$. Note that this separation of the antenna power pattern and the square of the waveform is possible because the mixed terms vanish in the quadratic form of Eq.~(\ref{eq2}) (see Eq.~(4) in \cite{Schutz2011}). We also assume that the one-sided PSD of the $k$-th detector's noise ($S_k(f)$) can be approximated with the detector's design PSD ($\widetilde{S}_{k}(f)$), obtained from \cite{InstWhitePaper2015,LCGT2009}, rescaled as $S_k(f) \approx (\widetilde{D}_{k} / D_{k})^2 \cdot \widetilde{S}_{k}(f)$, where $\widetilde{D}_{k}$ is the binary neutron star (BNS) range calculated for the design operation of the $k$-th detector, while $D_{k}$ is the expected BNS range in a given year for the same $k$-th detector (see Table 1).
\begin{table}[h!]
\centering
\begin{tabular}{lcccccccc}
\hline
 \multicolumn{1}{c}{GW detector}   & 2017 & 2018 & 2019 & 2020  \\ 
\hline
aLIGO-Hanford  & 120 & 170 & 200 & 200   \\ 
\hdashline
aLIGO-Livingston  & 120 & 170 & 200 & 200 \\ 
\hdashline
AdV   & 60 & 85 & 115 & 115 &  \\ 
\hdashline
KAGRA       &   &  & 259 & 259 &  \\ 
\hline 
\end{tabular}
\caption{\label{table1} Expected BNS ranges of different GW detectors in years between 2017-2020, given in megaparsec units \cite{Abbott2016}. The BNS range is the volume- and orientation-averaged distance at which the coalescence of two 1.4 $M_\odot$ neutron stars gives a matched filter SNR of 8 in a single detector \cite{Abbott2016}. As up-to-date BNS range estimates are currently unavailable for KAGRA, we use the ones published in \cite{LCGT2009}.}
\end{table}
Furthermore, in cases of characterizing network observations for pulsars, we compare the rescaled PSDs of individual detectors only at the expected frequency of the GW emitted by the pulsar, i.e.\ $S_k(f_\mathrm{GW}$). In these cases, our FoM can be written as:
\begin{equation}
\label{eq4}
\frac{\overline{\rho}_{\mathcal{N}\mathrm{sub}}^{2}(t) }{\overline{\rho}_{\mathcal{N}}^{2}(t) } =  \frac{\sum^{N-1}_{k=1}  \left[ F_{+,k}^{2}(t) + F_{\times,k}^{2}(t) \right] \sigma_k }{ \sum^{N}_{k=1} \left[ F_{+,k}^{2}(t) + F_{\times,k}^{2}(t) \right] \sigma_k } \, ,
\end{equation}
where $\sigma_k =  \widetilde{S}(f_\mathrm{GW})/S_k(f_\mathrm{GW})$ is a weighting factor that differentiates individual detectors' contribution to the network sensitivity according to their PSDs. In this definition, $\widetilde{S}(f_\mathrm{GW})$ is the design PSD of the most sensitive detector of the network at $f_\mathrm{GW}$. We note that the integral term of Eq.~(\ref{eq3}) appears both in the numerator and in the denominator in Eq.~(\ref{eq4}) and thus can be cancelled. This method allows us to take into consideration the differences of individual detector sensitivities in a simple but effective way which does not increase the computational time of simulations significantly. 

By choosing different $\mathcal{N}_{\mathrm{sub}}$ subnetworks from the same $\mathcal{N}$ network, and calculating the ratios of SNRs as functions of time using Eq.~(\ref{eq4}), we can calculate the ratio that a subnetwork contributes to the total network SNR as a function of time. We can generalize Eq.~(\ref{eq4}) to multiple sources by summing over antenna patterns calculated towards different sources defined by a set of $\alpha$ sky directions:
\begin{equation}
\label{eq5}
\frac{\overline{\rho}_{\mathcal{N}\mathrm{sub}}^{2}(t)}{\overline{\rho}_{\mathcal{N}}^{2}(t)} =   \frac{\sum_{\alpha}  \nu_\alpha \left( \sum^{N-1}_{k=1} \left[ F_{+,k,\alpha}^{2}(t) + F_{\times,k,\alpha}^{2}(t) \right] \sigma_{k} \right) }{\sum_{\alpha}  \nu_\alpha \left( \sum^{N}_{k=1} \left[ F_{+,k,\alpha}^{2}(t) + F_{\times,k,\alpha}^{2}(t) \right]\sigma_{k} \right) } \, ,
\end{equation}
where $\nu_\alpha$ is weighting individual pulsars. In section 2.2, we weight different pulsars in a multiple-source analysis as $\nu_\alpha \propto f^4_{\alpha}/ D_\alpha^2$, where $f_{\alpha}$ is the expected GW frequency, and $D_\alpha$ is the distance of the $\alpha$-th pulsar (see Table \ref{table2}) \cite{Ming2016}.

In section \ref{sec21} and \ref{sec22}, we examine only one realistic set of detectors as an $N$-detector network in order to demonstrate our method. This network consists of aLIGO-Hanford (H), aLIGO-Livingston (L) and AdV (V), and is expected to be operational from 2017; see Table \ref{table1}. We denote this set of detectors as $\mathcal{N} = \left\lbrace \mathrm{H,L,V} \right\rbrace $, and the sets of the 2-detector subnetworks as $\mathcal{N}_{\mathrm{sub}} = \left\lbrace \mathrm{H,L} \right\rbrace \mathrm{or} \left\lbrace \mathrm{H,V} \right\rbrace \mathrm{or} \left\lbrace \mathrm{L,V} \right\rbrace$.  In section \ref{sec23}, we extend our $N$-detector network to $ \mathcal{N} = \left\lbrace \mathrm{H,L,V,K} \right\rbrace $, where K stands for KAGRA. This 4-detector network is expected to be operational from 2019; see Table 1.

\subsection{Optimizing network operations for GW searches from the Crab and Vela pulsar}
\label{sec21}

Pulsars are promising candidates as sources of continuous GWs observable within the era of advanced GW detectors \cite{WhitePaper2014,Aasi2014}. Hence, we carried out two independent analyses with pulsars which have the highest known spin-down limits, namely the Crab pulsar \cite{Aasi2014} and Vela pulsar \cite{Aasi2014}. In these analyses, we calculate our FoMs towards the sky directions of these pulsars separately, and investigate, how FoMs change with time within a period of a sidereal day.
\begin{figure}[h!!]
    \centering
\begin{tabular}{cc}
    \hspace*{-10pt} \subfloat{\includegraphics[width=75mm,height=55mm]{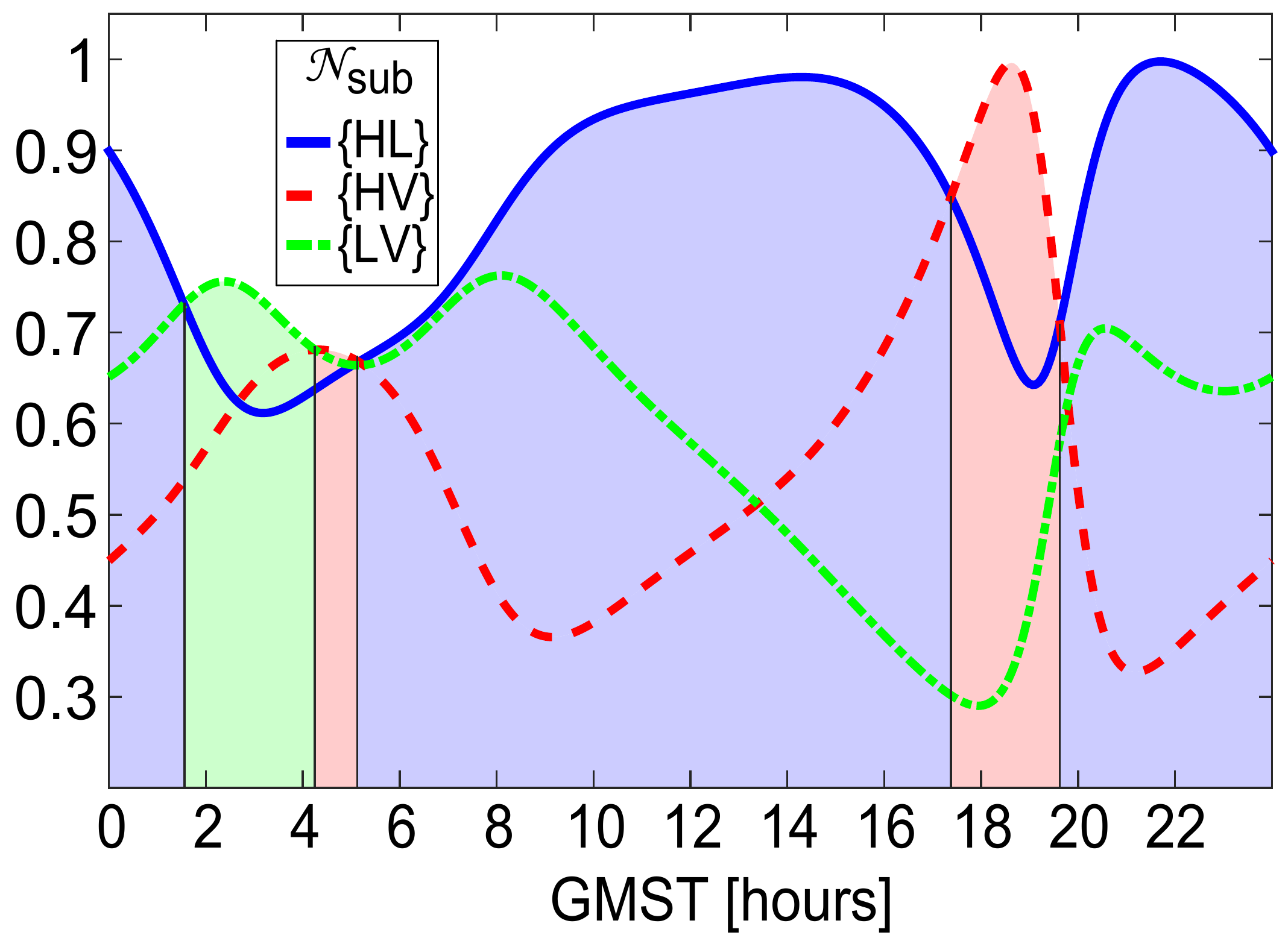}} \hspace*{-4pt} \subfloat{\includegraphics[width=80mm,height=55mm]{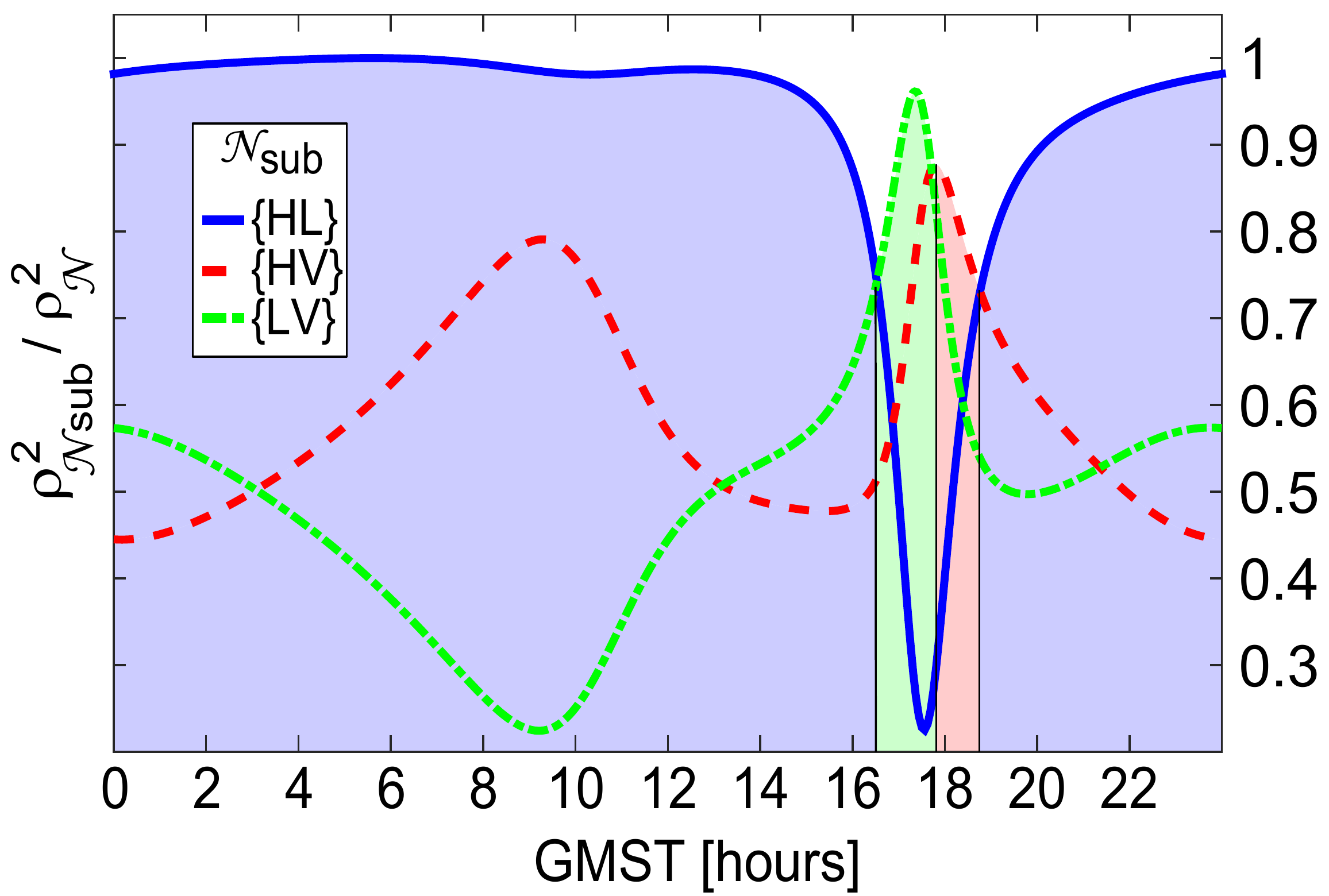}} \\
\end{tabular}
\caption{\label{fig1}Relative contributions of different 2-detector subnetworks to the total network SNR considering targeted searches of continuous GWs from the Crab pulsar (left panel) and the Vela pulsar (right panel). Directional sensitivities are expressed as ratios of the square of network SNRs collected within $\Delta t$ time windows around the given GMST satisfying $T_\mathrm{GW} \ll \Delta t\ll T_\mathrm{Earth}$. Here $\rho_{\mathcal{N}}$ is the SNR collected by the $\left\lbrace \mathrm{H,L,V} \right\rbrace$-network, and $\rho_{\mathcal{N}\mathrm{sub}}$ denotes the SNR collected by 2-detector subnetworks, where H stands for aLIGO-Hanford, L for aLIGO-Livingston, and V for Advanced Virgo. The colored regions indicate time intervals when different subnetworks (symbolized by the blue, red, and green colors) are the most sensitive ones compared to the others. Note that FoM values shown here have a daily periodicity due to Earth's rotation.}
\end{figure}
The results are shown in Fig.~\ref{fig1}, where the colored curves represent values of the FoM for each 2-detector subnetworks as a function of time. Colored regions below the envelopes indicate which subnetwork provides the highest FoM  temporally (i.e.\ has the highest contribution to the network SNR). 

We summarize our conlusions drawn from Fig.~\ref{fig1} in the following items:
\begin{itemize} 

\item In the search for continuous GWs from the Crab pulsar, the $\left\lbrace \mathrm{H,L} \right\rbrace$ subnetwork contributes the most to the network SNR during the day, except for a $\sim\!3$ hour long time window around $3$:$00$ Greenwich mean sidereal time (GMST) and a $\sim\!2$ hour long time window around $19$:$00$ GMST, when first the $\left\lbrace \mathrm{L,V} \right\rbrace$ subnetwork, and then the $\left\lbrace \mathrm{H,V} \right\rbrace$ subnetwork has higher contribution to the network SNR than the other two subnetwork configurations. Thus, in order to minimize the SNR loss for the network due to e.g.\ service operations resulting with elevated noise levels or temporal lock losses, (if possible) one should try to schedule such operations at aLIGO-Livingston to around $19$:$00$ GMST, or at aLIGO-Hanford around $3$:$00$ GMST, during a day. From these, we conclude that aLIGO-Hanford is the key contributor to this scientific goal, because subnetworks containing aLIGO-Hanford have the highest contribution to the network SNR (e.g.\ a temporal lock loss of aLIGO-Hanford around $18$:$00$ GMST causes more than $70\%$ loss in the FoM compared to the 3-detector case).

\item Results for the continuous GW search from the Vela pulsar reveal that the $\left\lbrace \mathrm{H,L} \right\rbrace$ subnetwork has also the highest contribution to the network SNR during almost the whole day, however, there is a $\sim\!2$ hour long time interval around $18$:$00$ GMST when $\left\lbrace \mathrm{H,L} \right\rbrace$ subnetwork's FoM decreases by more than $70\%$. As both the $\left\lbrace \mathrm{H,V} \right\rbrace$ and $\left\lbrace \mathrm{L,V} \right\rbrace$ subnetworks have almost equally relevant contribution to the network SNR around $18$:$00$ GMST, it is the AdV detector that contributes the most to the network SNR within this time window of the day. Thus a short-term service operation is suggested to be timed around $17$:$00$ GMST for aLIGO-Hanford, and around $18$:$00$ GMST for aLIGO-Livingston. As an example, a narrow-band search of a continuous signal with the two LIGO detectors (i.e.\ detectors tuned to the expected frequency band of the Vela pulsar's GW emission) should not be scheduled between $16$:$00$ GMST and $20$:$00$ GMST, because of the high level of SNR loss within this period. According to our results, aLIGO-Hanford and aLIGO-Livingston are the key contributors to the scientific goal.

\end{itemize}
We note that these results were calculated by the expected operation sensitivities for both LIGO and AdV detectors in 2017 (see Table \ref{table1} for details). Even though our results are provided using theoretical sensitivity curves of detectors that do not change with time, short-term prompt changes in the detector sensitivities can also be taken into account, and the FoM can be recalculated using real-time data of detectors' sensitivity during observing runs.

\subsection{Optimizing network operations for GW searches from 10 high interest pulsars}
\label{sec22}

In this section, we present results on multiple-source targeted characterization of networks towards 10 high interest pulsars, which are the ones known to have the smallest distances and the highest spin-down rates \cite{WhitePaper2014, Aasi2014}. These pulsars, namely the Crab, Vela, J1833-1034, J1952+3252, J0537-6910, J0205+6449, J1813-1749, J1913+1011, J0737-3039A, J1813-1246, were marked out as the most promising candidates for a GW detection in the advanced detector era, (see Table \ref{table2}) \cite{WhitePaper2014,Aasi2014}.
\begin{table}[]
\centering
\begin{tabular}{lr}
\subfloat{\begin{tabular}{ccc}
\hline
             & $f_\mathrm{GW}$ {[}Hz{]}  & $D$ {[}kpc{]} \\ \hline
Crab         & $59.44$                    & $2.0$           \\ \hdashline
Vela         & $22.39$                    & $0.29$        \\ \hdashline
$J1833-1034$ & $32.33$                    & $4.8$           \\ \hdashline
$J1952+3252$ & $50.59$                   & $3.0$           \\ \hdashline
$J0537-6910$ & $123.94$                  & $50.0$         \\ \hline
\end{tabular}}
\hspace*{20pt}
\subfloat{\begin{tabular}{ccc}
\hline
              & $f_\mathrm{GW}$ {[}Hz{]} & $D$ {[}kpc{]}    \\ \hline
$J0205+6449$  & $30.45^a$         & $3.2^b$ \\ \hdashline
$J1813-1749$  &  $44.74^a$         &  $4.8^c$ \\ \hdashline
$J1913+1011$  & $55.70$         & $4.5$ \\ \hdashline
$J0737-3039A$ & $88.11$         & $1.1$ \\ \hdashline
$J1813-1246$  & $41.60$         & $1.9$ \\ \hline
\end{tabular}}
\end{tabular}
\caption{\label{table2}Expected frequencies of potential continuous GWs ($f_\mathrm{GW}$) from 10 selected pulsar sources with their distances ($D$) \cite{Aasi2014}, $^a$\cite{WhitePaper2014}, $^b$\cite{Abdo2009}, $^c$\cite{Halpern2012}.}
\end{table}
 
\begin{figure}[h!]
    \centering
	{\includegraphics[width=115mm, height=65mm]{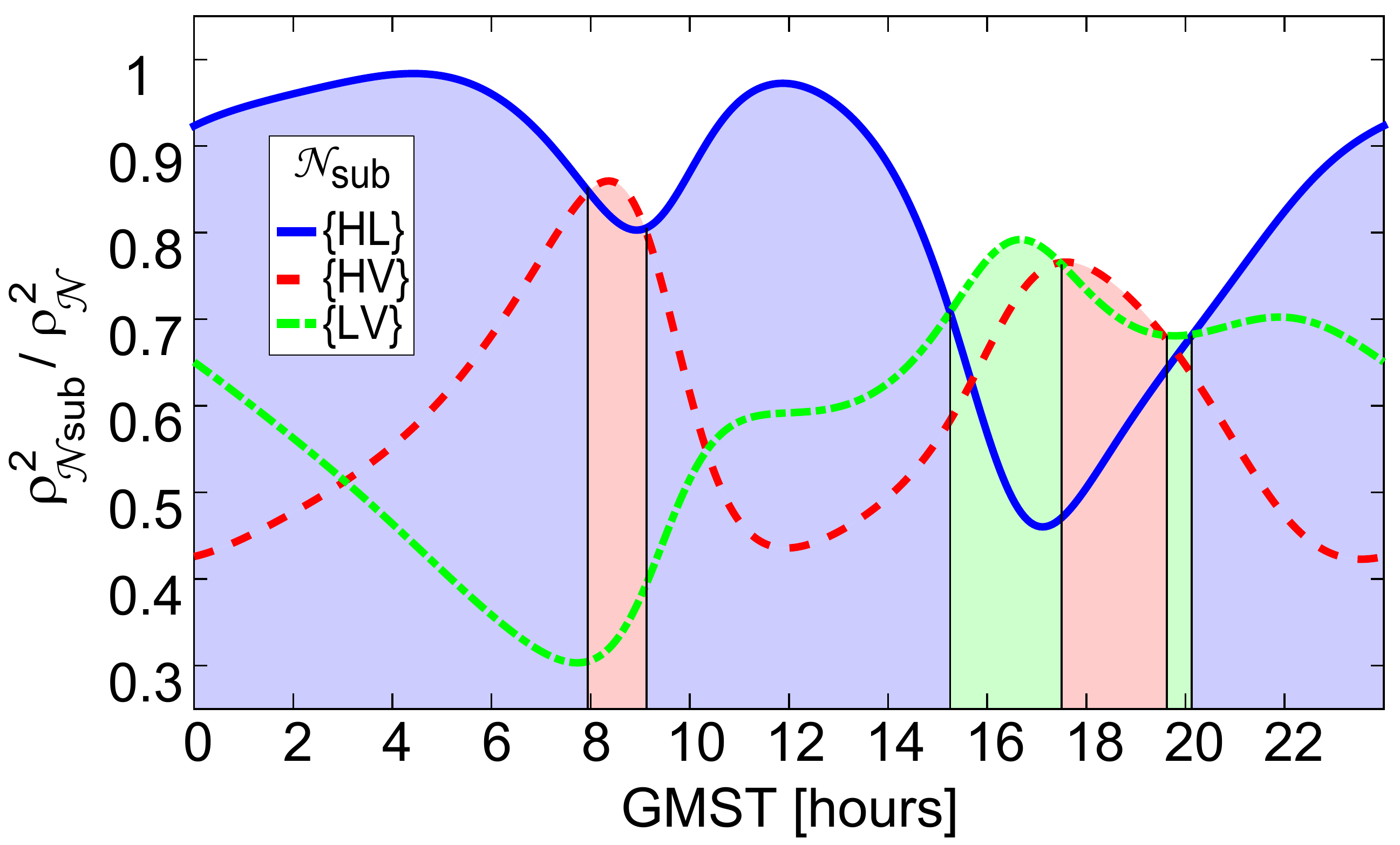}}  \\
\caption{\label{fig2}Relative contributions of different 2-detector subnetworks to the total 3-detector network SNR considering targeted searches of GWs from 10 high interest pulsars simultaneously in function of time. Notations of the plot are the same as the ones in Fig.~\ref{fig1}.}
\end{figure}

We carried out a multiple-source analysis by using Eq.~(\ref{eq5}) with $\alpha \in \left[ 1,10 \right]$. Results are shown in Fig.~\ref{fig2}, where it is revealed that the $\left\lbrace \mathrm{H,L} \right\rbrace$ subnetwork has the highest contribution to the network SNR, except for $\sim\!1$ hour long interval near $8$:$00$ GMST, and a $\sim\!5$ hour long time window from $15$:$00$ GMST to $20$:$00$ GMST, when first the $\left\lbrace \mathrm{H,V} \right\rbrace$ subnetwork, then either the $\left\lbrace \mathrm{L,V} \right\rbrace$ or the $\left\lbrace \mathrm{H,V} \right\rbrace$ subnetworks have higher contribution to the network SNR. Therefore, we consider both aLIGO-Hanford and aLIGO-Livingston as being the key contributors of this scientific goal. The expected operational sensitivities of detectors were implemented in this analysis by using BNS ranges predicted for network operations in 2017. In this analysis, we weighted each pulsar with their know frequencies and distances according to the weighting factor $\nu$ introduced in Eq.~(\ref{eq5}).

\subsection{Optimizing network operations for GW searches from the Galactic Center}
\label{sec23}

After demonstrating the capability of the method for cases of individual sky locations, as well as for multiple sky locations, we apply the method for a whole sky region. Here the goal is to characterize network operations with respect to searches for continuous GWs from the Galactic Center, which is defined as a circular region around the center of the Milky Way with a radius of $200$ parsecs  \cite{Aasi2013b} (covering $\sim 7$ square degrees, or $\sim 0.02$ percent of the sky). Technically, we split this circular region into $1582$ discrete pixels of equal solid angles, and we treated each pixels as point sources with equal weights, located in the geometrical center of pixels (see our calculations with point-like sources in section \ref{sec22}). This region of the sky is presumably dense with massive compact objects \cite{Aasi2013b} such as rapidly rotating neutron stars, which are potential sources of continuous GWs. Motivated by this expectation, we carried out two analyses with two different sets of detectors. One set is the previously introduced $\mathcal{N} = \left\lbrace \mathrm{H,L,V} \right\rbrace $ network, while the other is the same network extended with KAGRA (K), i.e.\ $\mathcal{N}= \left\lbrace \mathrm{H,L,V,K} \right\rbrace $. In this latter case the possible sets of 3-detector subnetworks are 
$\mathcal{N}_{\mathrm{sub}}$ = $\left\lbrace \mathrm{H,L,V} \right\rbrace$, $\left\lbrace \mathrm{H,L,K} \right\rbrace$, $\left\lbrace \mathrm{H,V,K} \right\rbrace$ and $\left\lbrace \mathrm{L,V,K} \right\rbrace$. In these analyses, we assume that compact objects are homogeneously distributed within the circular region around the Galactic Center; applying specific formation and distribution models could increase the accuracy of our method, however, it is beyond the scope of this paper. 
\begin{figure}[h!]
    \centering
\begin{tabular}{cc}
    \hspace*{-10pt} \subfloat{\includegraphics[width=73mm,height=55mm]
    {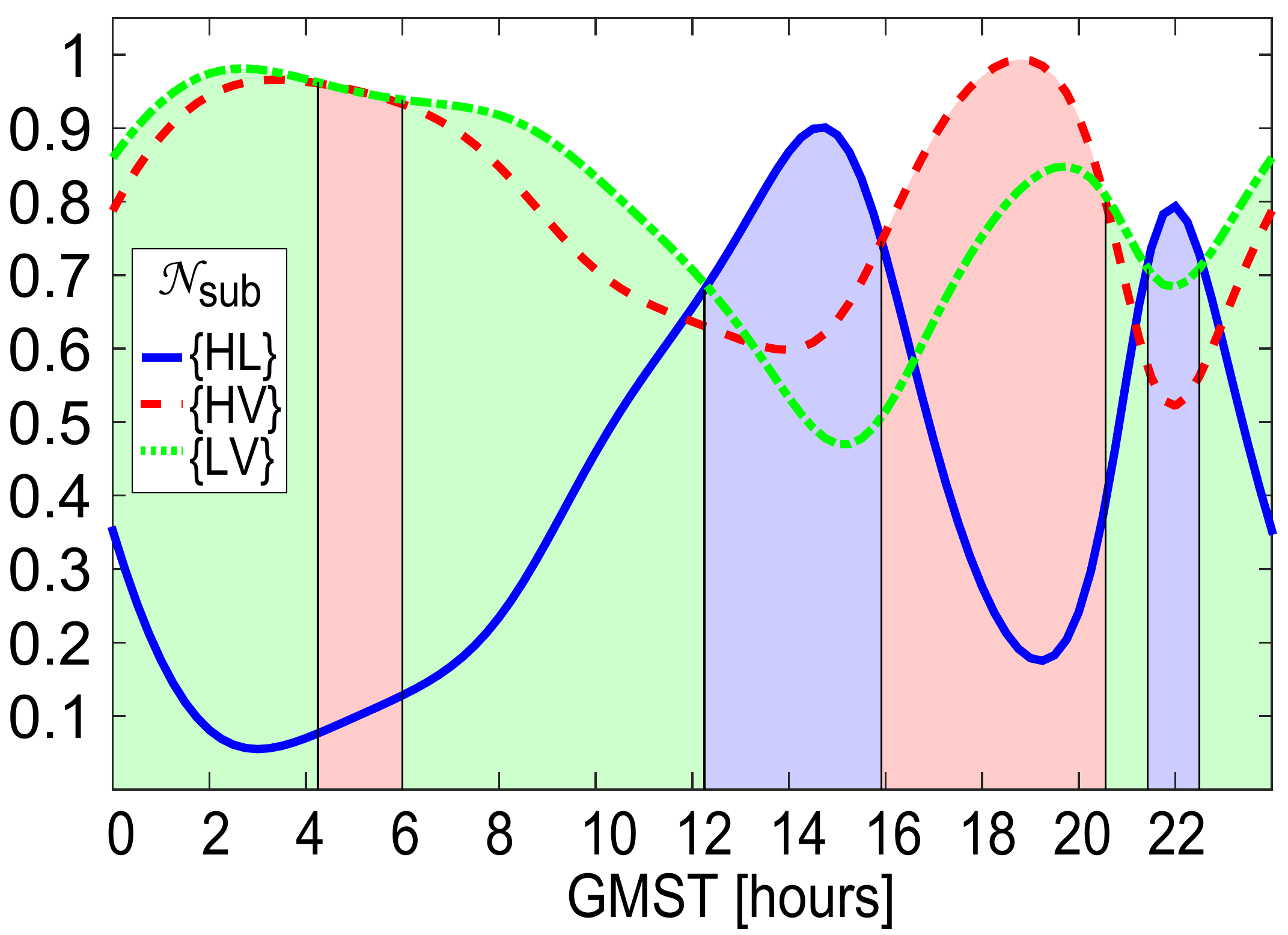}} \hspace*{-4pt}
    \subfloat
{\includegraphics[width=80mm,height=55mm]
    {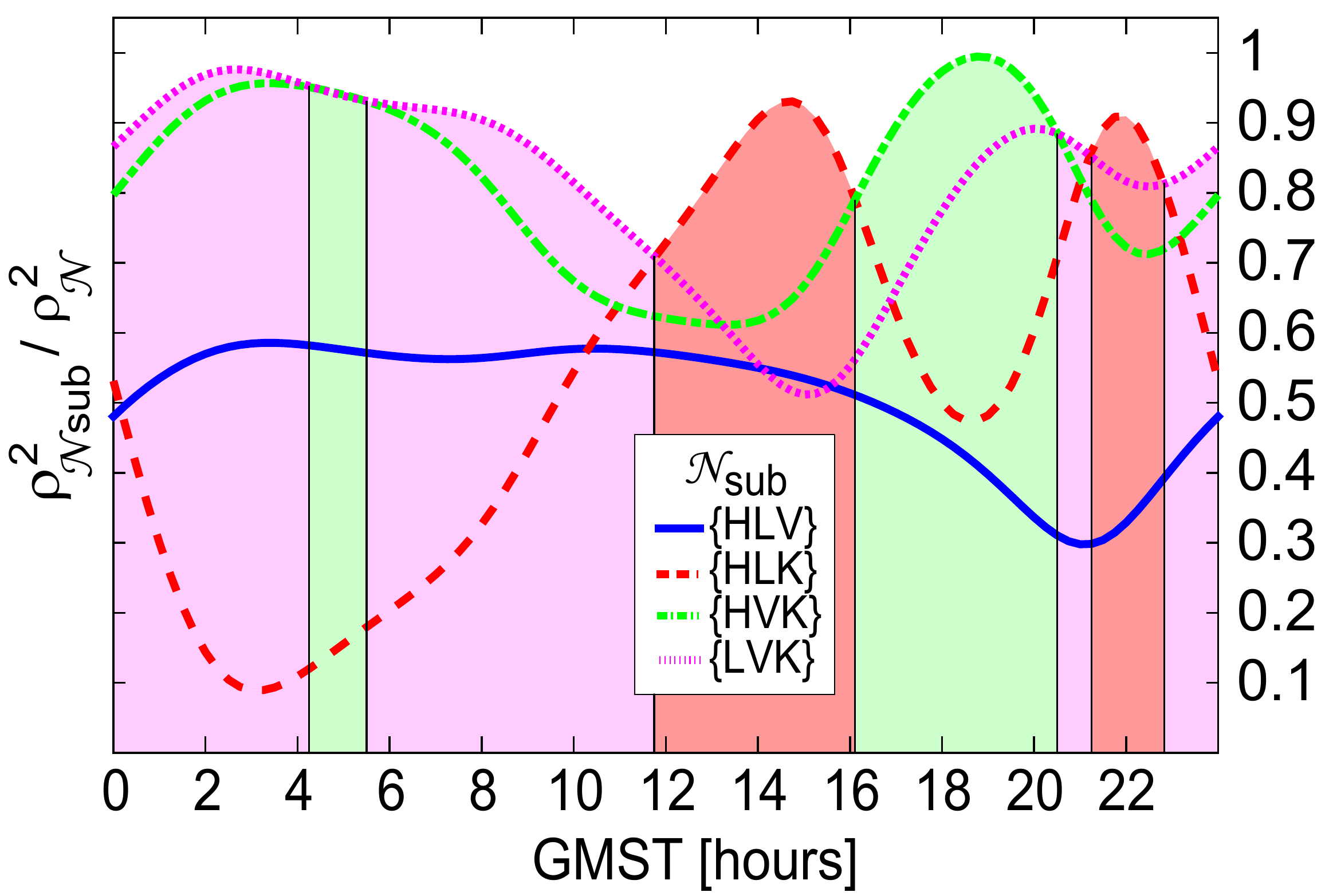}} \\
\end{tabular}
\caption{\label{fig3}Relative contributions of different 2-detector subnetworks of the 2017 $\mathcal{N} = \left\lbrace \mathrm{H,L,V} \right\rbrace $ network (left panel), and 3-detector subnetworks of the 2019-2020 \ $\mathcal{N}= \left\lbrace \mathrm{H,L,V,K} \right\rbrace $ (right panel) to total network SNRs considering targeted searches of GWs from the Galactic Center region. Notations of the plot are the same as the ones in Fig.~\ref{fig1}.}
\end{figure}
Results are shown in Fig.~\ref{fig3} for the $3$-detector network in the left panel, and for the 4-detector network in the right panel. Here, notations of axes, curves and colored regions are the same as in Fig.~\ref{fig1}. We summarize our findings based on Fig.~\ref{fig3} as follows:
\begin{itemize}
\item In case of analyzing results obtained for the 3-detector network (left panel), we cannot determine a key contributor for the whole day, but we can identify temporal key contributors indicated by colors. Each time intervals corresponding to these colored regions can provide the basis for the scheduling of short-term services, i.e.\ the green region indicates an optimal time interval for timing the service operation of aLIGO-Hanford, red is the optimal period for aLIGO-Livingston, and the blue is optimal for AdV, respectively.

\item In case of the 4-detector network (right panel), the most remarkable drop (approximately than $90\%$) in the FoM value appears for the $\left\lbrace \mathrm{H,L,K} \right\rbrace$ subnetwork around $3$:$00$ GMST, where it is AdV that contributes the most to the FoM. Other minimums indicate other key contributors, and the colored regions correspond to the most relevant subnetworks in terms of their contribution to the network SNR. Time intervals corresponding to these colored regions can provide the basis for the scheduling of short-term services, i.e.\ the green region indicates an optimal time interval for timing the service operation of aLIGO-Livingston, purple is an optimal period for aLIGO-Hanford, and the red is optimal for KAGRA, respectively.
\end{itemize}
The operational sensitivities were approximated in these analyses by using the expected BNS ranges of 2017 for the 3-detector network case, and of 2019 for the 4-detector network case, see Table \ref{table1}.

\section{Characterizing network operations with respect to joint GW+EM detections of neutron star binary coalescences}
\label{sec3}

In this section, we introduce the FoM we use to characterize detector networks with respect to the rate of GW triggers that can provide for joint GW+EM detections of neutron star binary coalescence events (for a conceptual overview, see \cite{WhitePaper2014} and references therein). The coalescence of a compact binary system consisting of either (i) two neutron stars (from now on, denoted by NSNS) or (ii) a neutron star and a black hole (from now on, denoted by NSBH), is a promising candidate as a source of nearly coincident GW and EM signals \cite{Kanner2012,Bartos2012,Metzger2011}. Because of this, triggering EM follow-up observations was marked out by the LIGO-Virgo Collaboration as a high priority scientific goal in the era of advanced GW detectors \cite{WhitePaper2014}. 

In order to accomplish a successful observation of an EM counterpart of a GW signal, an effective sky localization of the GW source is crucial. Using at least 3 detectors, the expected size of the localization area is $\sim\!10 - 100$ square degrees \cite{Hanna2014}, which makes the identification of EM counterparts of GW triggers challenging, even with wide-field telescopes. However, we know from both binary formation models \cite{Abadie2010,Phinney1991,Postnov2014,Kalogera2007} and from observations of short gamma-ray bursts \cite{Fong2010} that NSNS and NSBH binaries undergo coalescences inside or in the near proximity of galaxies. Thus, targeting follow-up observations to sky positions of known galaxies within the 3-dimensional error box of the event reconstructed by the GW detector network, can reduce the sky area to be observed by a factor of $\mathcal{O}(10^3)$ \cite{Bartos2015}, and thereby can greatly enhance the probability for the successful detection of an EM counterpart \cite{Hanna2014,Abadie2012}. As it has been shown in multiple papers (see e.g.\ \cite{Hanna2014,Abadie2012}), following this observing strategy is beneficial for the hunting for EM counterparts, even if the applied catalog of local galaxies is incomplete.

Following this logic, we have made the simplifying assumption in our analysis that a joint GW+EM detection of a neutron star binary coalescence can only be achieved, if the corresponding GW trigger sent out by the GW detector network to EM observing partners originates from a galaxy that is included in the galaxy catalog used for EM target selections. Even though it is possible to use an EM observing strategy different from the one outlined here, as well as there is a possibility for a joint GW+EM detection even if targeting is not based on a galaxy catalog of limited size, we focus our analysis on this well-known case.

From a set of existing galaxy catalogs \cite{Igor2013,Freedman2001,Mateo1998,Patruel2003,Kopparapu2008}, we chose the \textit{Gravitational Wave Galaxy Catalogue} (GWGC) \cite{White2011} for our analysis, because the GWGC is applied for target selections by EM follow-up partners of the LIGO-Virgo Collaboration \cite{Sopuerta2015}. Note that there are ongoing efforts to construct larger and more complete full-sky catalogs dedicated to EM follow-up efforts in the advanced GW detector era. Therefore we only use the GWGC to demonstrate how our analysis works, and propose to use more up-to-date catalogs in the future as they become available.

By applying the GWGC, we aimed to predict the number of GW triggers that can be expected from NSNS + NSBH coalescences in association with galaxies listed in the catalog. With this, we define our FoM as the total rate of NSNS + NSBH detections by a GW detector network from all GWGC galaxies. Note that all such NSNS + NSBH coalescence events would trigger EM follow-up observations if the GW detectors operate with sensitivities outlined in Table \ref{table1}. The rates are calculated within time windows with arbitrary durations of $\Delta t \ll T_\mathrm{Earth}$ around given $t$ times of a sidereal day, and thus are given as functions of $t$ ($T_\mathrm{Earth}$ being Earth's rotational period). We give the formal definition of the FoM, denoted as $\widetilde{\mathcal{R}}(t)$, as:
\begin{equation}
\label{eq6}
\widetilde{\mathcal{R}}(t) = \sum_{n =1}^{N_{\mathrm{gal}}} \left(  \widetilde{\mathcal{R}}_{n,\mathrm{NSNS}}(t) + \widetilde{\mathcal{R}}_{n,\mathrm{NSBH}}(t) \right) \, ,
\end{equation}
where the sum goes over the number of galaxies in the catalog, while $\widetilde{\mathcal{R}}_{n,\mathrm{NSNS}}(t)$ and $\widetilde{\mathcal{R}}_{n,\mathrm{NSBH}}(t)$ are the estimated GW detection rates for NSNS and NSBH coalescences, respectively, assigned to individual galaxies. Detection rates of NSNS and NSBH binaries are calculated from merger rates associated to host galaxies as follows: 
\begin{eqnarray} \label{eq7}
\widetilde{\mathcal{R}}_{n}(t)  = \int_{\mathrm{min}(m_1)}^{\mathrm{max}(m_1)} \int_{\mathrm{min}(m_2)}^{\mathrm{max}(m_2)} \mathcal{P}(m_1,m_2) \, \mathcal{R}_{n} \times & \\
           \nonumber && \hskip -1in
  H(\rho_{n}(t,m_1,m_2)-\rho_\mathrm{lim}) \, \mathrm{d} m_1 \mathrm{d} m_2 \, ,
\end{eqnarray}
where $\mathcal{P}(m_1,m_2)$ is the normalized distribution of component masses in the NSNS or NSBH system residing in a Milky Way Equivalent Galaxy (MWEG). $\mathcal{R}_{n}$ denotes the total merger rate of a binary type assigned to the $n$-th galaxy. $H(x)$ is the Heaviside step function, in which $\rho_n(t,m_1,m_2)$ denotes the SNR of a GW detected at time $t$ of a binary residing in the $n$-th galaxy with component masses $m_1$ and $m_2$; and $\rho_\mathrm{lim}$ denotes the SNR threshold for claiming a GW detection (conventionally, $\rho_\mathrm{lim}=8$). Throughout in our analysis, we use \textit{model C} introduced in \cite{Belczynski2007} for calculating $\mathcal{P}(m_1,m_2)$ for both binary types, with component mass ranges of $0.9 - 10.8 \; M_\odot$ for NSBH binaries (with mass ratio $\sim 0.1-0.4$) and $1.15-2.5 \; M_\odot$ for NSNS binaries (with mass ratio $>0.9$) (see distributions in Figure 6. in \cite{Belczynski2007}).

We calculate $\rho_{n}(t,m_1,m_2)$ using Eq.~(\ref{eq1}), where we redefine the limits of the frequency integration (the new lower limit is the minimum frequency at which the GW enters the sensitive band of the detector, the new upper limit is the frequency that corresponds to the innermost stable circular orbit (ISCO), i.e. $f_\mathrm{ISCO}=c^3/(6^{3/2} \pi G M)$, where $M$ is the redshifted total mass of the binary \cite{Cutler1994}); and also replace the continuous GW spectrum to the 1.5 post-Newtonian spectrum (here denoted as $\widetilde{h}_+(f, m_1, m_2)$ and $\widetilde{h}_{\times}(f, m_1, m_2)$) of a spinless circular inspiraling binary with component masses $m_1$ and $m_2$ as \cite{Cutler1994}:
\begin{equation} \label{eq9}
  \widetilde{h}_{k,n}(f, t, m_1, m_2) = F_{+,k,n}(t) \widetilde{h}_+(f, m_1, m_2) + F_{\times,k,n}(t) \widetilde{h}_{\times}(f, m_1, m_2) \, ,
\end{equation}
where $k$ refers to the $k$-th detector, while $n$ refers to the $n$-th galaxy in the catalog. In this model, we utilize that both $F_+(t)$ and $F_{\times}(t)$ are varying slowly with time compared to durations of binary merger signals in the detectors' sensitive band, and thus we treat them as being constants throughout the integration of the signal. We emphasize that both $\widetilde{h}_+(f)$ and $\widetilde{h}_{\times}(f)$ are proportional to the reciprocal of the host's distance from Earth \cite{Cutler1994}. This also means that there is a distance limit corresponding to the $\rho_{n} > \rho_\mathrm{lim}$ criterion set by Eq.~(\ref{eq7}) which determines whether a GW can be detected or not from the given host. Polarization angles of binaries are assumed to be random and isotropically distributed over the range of $[0, 2\pi]$, therefore we average detector SNRs over polarization angles, the same way as in Eq.~(\ref{eq3}). 

Following \cite{Abadie2010}, we approximate $\mathcal{R}_{n}$ in Eq.~(\ref{eq7}) as:
\begin{equation}
\label{eq10}
\mathcal{R}_{n} = \frac{L_{n}}{L_\mathrm{MW}} \mathcal{R}_{\mathrm{MWEG}} \, ,
\end{equation}
where $L_{n}$ denotes the blue-light luminosity of the $n$-th galaxy in the catalog, $L_\mathrm{MW} = 1.7 \times 10^{10} L_{B,\odot}$ denotes the blue-light luminosity of the Milky Way \cite{Abadie2010}, and $\mathcal{R}_{\mathrm{MWEG}}$ is the merger rate of the corresponding binary type assigned to a MWEG. We obtain $\mathcal{R}_{\mathrm{MWEG}}$ values by calculating the median of \textit{model C} data given in Table 2.~of \cite{Belczynski2007}. In Fig. 4, we shown $\mathcal{R}_{n}$ values we calculated using Eq.~(\ref{eq10}) for the 49 183 GWGC galaxies that have measured B-band magnitudes.
\begin{figure}[h!!]
    \centering
\includegraphics[width=140mm]{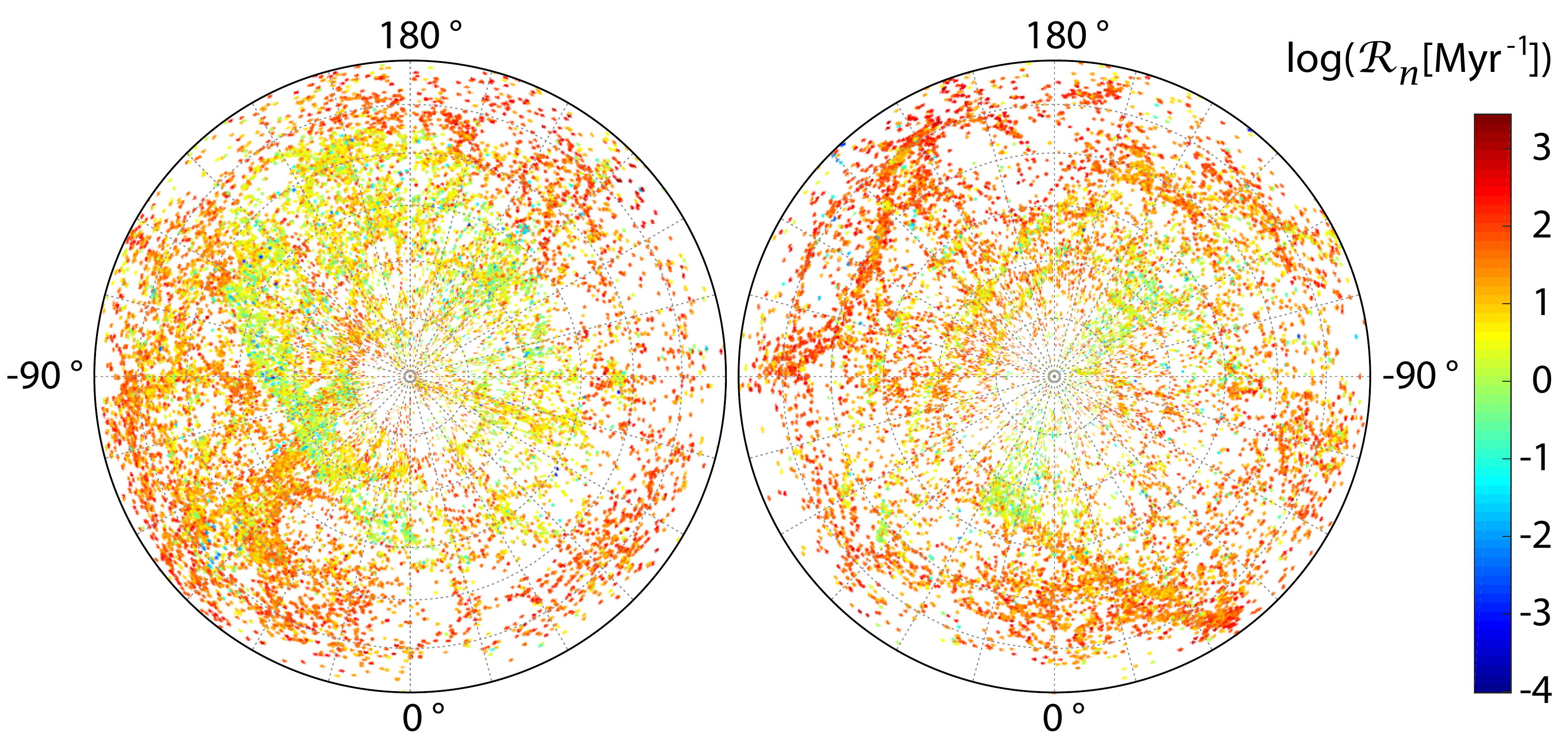}
\caption{\label{fig4}The sky map of GWGC galaxies where the coloring represents the estimated rate of NSBH+NSNS binary mergers within each galaxy. The rate is given in units of mergers per million years, and the sky map is given in galactic coordinates that scale from -180$^\circ$ to 180$^\circ$ counterclockwise on the left panel (northern galactic hemisphere), and clockwise on the right panel (southern galactic hemisphere).}
\end{figure}

\begin{figure}[h!!]
    \centering
\includegraphics[width=130mm]{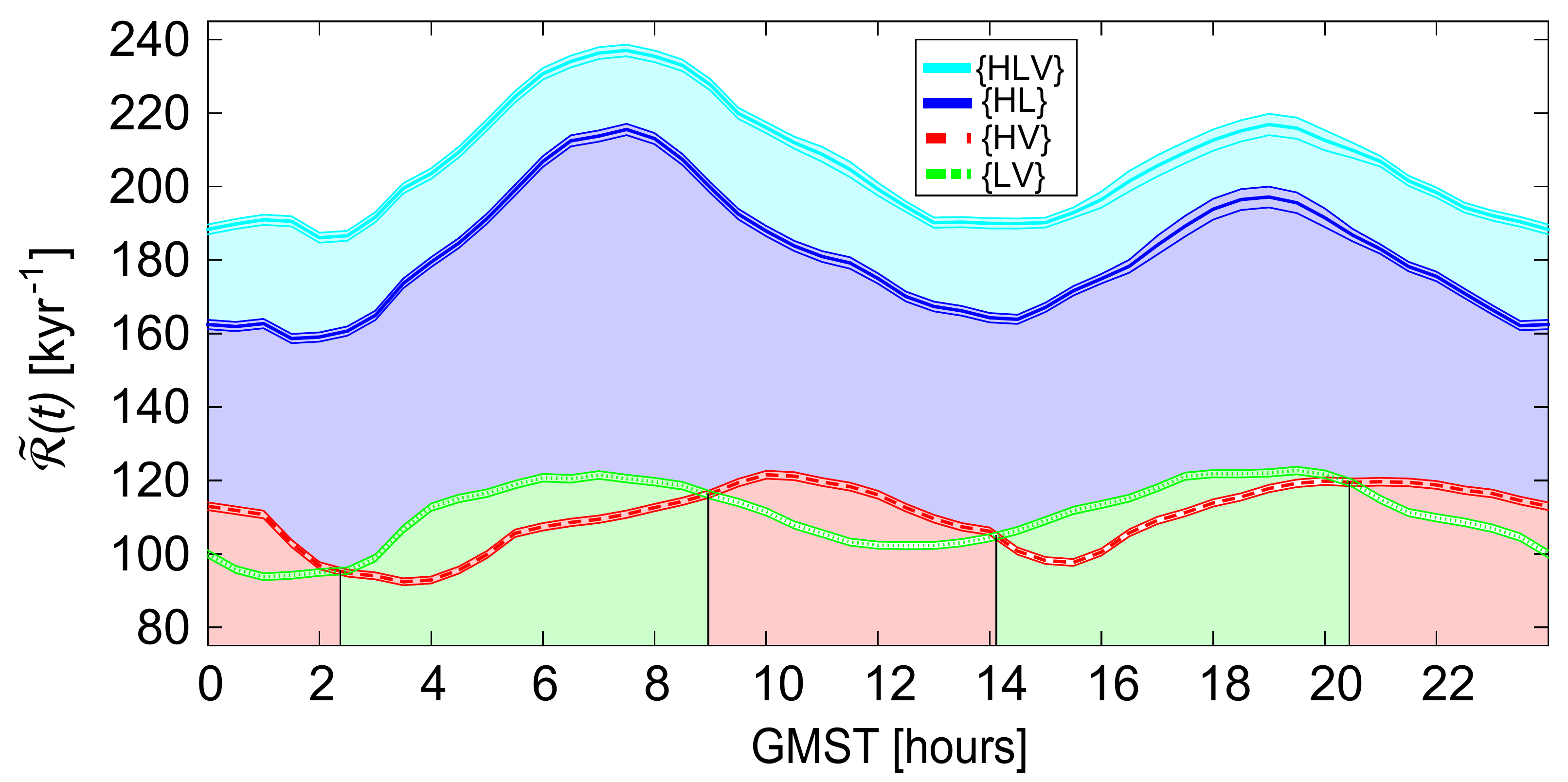}
\caption{\label{fig5}Detection rates of GW triggers from NSNS+NSBH coalescence events as the function of time within a sidereal day in 2018 with expected sensitivities of Table 1. Detectors considered in this analysis are the aLIGO-Hanford (H), aLIGO-Livingston (L), AdV (V). Colored regions indicate the time intervals when given subnetworks (represented by the colors) provide the highest trigger rate.}
\end{figure}

Following the mathematical formulation we have defined so far, we present our results for the $\left\lbrace \mathrm{H,L,V} \right\rbrace$ network, as well as all for its 2-detector subnetworks, with respect to detecting GWs of neutron star binary coalescences from GWGC hosts. By knowing $\mathcal{R}_{n}$ for all GWGC galaxies, as well as the network parameters, we are able to estimate the detection rates of GW triggers for each GWGC galaxy as a function of time using Eq.~(\ref{eq7}). By summing over all galaxies, we obtain our FoM for the network and subnetworks using Eq.~(\ref{eq6}). In Fig.~\ref{fig5}, we present $\widetilde{\mathcal{R}}(t)$ in kyr$^{-1}$ units as a function of time for a period of a sidereal day. According to results shown in Fig.~\ref{fig5}, the FoM for the $\left\lbrace \mathrm{H,L} \right\rbrace$ subnetwork always exceeds the FoM for 2-detector subnetworks including AdV. 
This is due to the fact that the expected strain sensitivities of aLIGO detectors are significantly higher than AdV's in 2018 (see Table \ref{table1}). The complete lock-loss of one aLIGO detector results with either the $\left\lbrace \mathrm{L,V} \right\rbrace$ subnetwork (denoted by the green curve in Fig.~\ref{fig5}) or the $\left\lbrace \mathrm{H,V} \right\rbrace$ subnetwork (denoted by the red curve in Fig.~\ref{fig5}) operating. In these cases, red regions indicate time intervals when the expected rate of GW detections from neutron star binaries in GWGC galaxies is higher for the $\left\lbrace \mathrm{H,V} \right\rbrace$ subnetwork, while the green regions indicate the same for the $\left\lbrace \mathrm{L,V} \right\rbrace$ subnetwork. We note that these results are mainly the effect of the large number of galaxies missing in the Galactic plane due to obscuration, which is an effect that is expected to limit EM counterpart observations of GW signals too.

\section{Conclusion}
\label{sec4}

In this paper, we have introduced two new time-dependent FoMs for characterizing GW detector networks and their subnetworks with respect to the scientific goals of (i) detecting continuous GWs from known sources, as well as of (ii) detecting signals from neutron star binary coalescences in known hosts. The FoM used with respect to the scientific goal (i) measures the ratio between the SNR collected by a network and the SNR collected by its subnetwork (for details, see section \ref{sec2}). The FoM used with respect to the scientific goal (ii) measures the rate of GW triggers from coalescing neutron star binaries that can provide for joint GW and EM detections (for details, see section \ref{sec3}).

We characterized the $\left\lbrace \mathrm{H,L,V} \right\rbrace$ network, and its 2-detector subnetworks by calculating the corresponding FoMs with respect to the scientific goal of detecting continuous GW from the Crab and Vela pulsar, respectively (in section \ref{sec21}). We presented these FoMs as a function of time in Fig.~\ref{fig1}. Using the FoM, we identified the aLIGO-Hanford as the key contributor to the goal of detecting GWs from the Crab pulsar, and both LIGO detectors as key contributors to the goal of detecting GWs from the Vela pulsar. We also suggested time periods to which service works resulting with elevated noise levels or lock losses for the detector should be scheduled in order to minimize the SNR loss of the whole network with respect to the given scientific goal. As an example, we have found that an $\lesssim$$2$ hour service operation of aLIGO-Livingston should be scheduled around $19$:$00$ GMST (if possible) when considering the goal of detecting a GW from the Crab pulsar, or around $17$:$00$ GMST for aLIGO-Hanford and around $18$:$00$ GMST for aLIGO-Livingston when considering the goal of detecting a GW from the Vela pulsar. According to our results, the $\left\lbrace \mathrm{H,L} \right\rbrace$ subnetwork contributes the most to achieving both scientific goals. We characterized the same networks as in section \ref{sec21} with respect to the scientific goal of simultaneous GW searches from $10$ high interest pulsars (in section \ref{sec22}). We have found that the $\left\lbrace \mathrm{H,L} \right\rbrace$ subnetwork contributes the most to achieving this goal compared to other subnetworks (see results in Fig.~\ref{fig2}). We carried out two analyses with respect to targeted GW search from the Galactic Center: one with $\left\lbrace \mathrm{H,L,V} \right\rbrace$ network, and another with $\left\lbrace \mathrm{H,L,V,K} \right\rbrace$ network (see details in section \ref{sec23}). In case of both the $\left\lbrace \mathrm{H,L,V} \right\rbrace$ and $\left\lbrace \mathrm{H,L,V,K} \right\rbrace$ networks, there are no individual key contributors, however, we identified time intervals within a sidereal day when potential service works for aLIGO-Hanford, aLIGO-Livingston, Advanced Virgo and KAGRA should be avoided.

We characterized the $\left\lbrace \mathrm{H,L,V} \right\rbrace$ network and its subnetworks in terms of detecting GWs from neutron star binary coalescences that hosted in galaxies of the local universe (in section \ref{sec3}). We calculated the detection rate of such events that have the potential for triggering EM follow-up observations using the GWGC galaxy catalog for identifying hosts (see results of the predicted detection rates as a function of time in Fig.~\ref{fig5}). Examining the results, we have found that the $\left\lbrace \mathrm{H,L} \right\rbrace$ subnetwork contributes the most to achieving this goal compared to other subnetworks. 

By identifying these key contributors, as well as time intervals, we are able to point out optimal operational strategies such as scheduling short-term services, when high noise activities around the key contributor detectors do not effect the observation; or by developing these methods to a real-time, target-based sensitivity monitoring tool of network operations as well. We can also suggest the presented methods for supporting efforts on developing techniques of prompt adjustments in detector sensitivities, e.g.\ switching detectors between broad-band and narrow-band operating modes \cite{InstWhitePaper2015,LCGT2009}. Based on these methods, an online monitoring tool can be developed which can be proposed to be installed in the detectors' control rooms.

\section{Acknowledgments}
\label{sec5}

This paper was reviewed by the LIGO Scientific Collaboration under LIGO Document P1600172. We would like to thank Imre Bartos, Yiming Hu, Szabolcs M\'arka, Zsuzsa M\'arka, Gergely M\'ath\'e, Tibor Rakovszky, and B\'alint T\'oth for their valuable comments on the manuscript. The authors are grateful for the support of the LIGO-Virgo Collaboration. This project has been supported by the Hungarian National Research, Development and Innovation Office $\--$ NKFIH K-115709. P\'eter Raffai is grateful for the support of the Hungarian Academy of Sciences through the "Bolyai J\'anos" Research Scholarship programme. Gergely D\'alya is supported through the New National Excellence Program of the Ministry of the Human Capacities.

\section*{References}
\bibliographystyle{unsrt}
\bibliography{References}

\begin{thebibliography}{10}

\bibitem{Abbott2015}
LIGO~Scientific Collaboration.
\newblock {\em Classical and Quantum Gravity}, 32:074001, 2015.

\bibitem{Acernese2015}
F.~Acernese et~al.
\newblock {\em Classical and Quantum Gravity}, 32:024001, 2015.

\bibitem{ADEOverview}
D.~Reitze.
\newblock {Advanced Detector Era Overview}.
\newblock {\em LIGO-G1300852-v5}, 2013.

\bibitem{Somiya2012}
K.~Somiya.
\newblock {\em Classical and Quantum Gravity}, 29:124007, 2012.

\bibitem{Iyer2011}
B.~Iyer; T. Souradeep; C. S. Unnikrishnan; S. Dhurandhar; S. Raja;~A. Sengupta.
\newblock {\em LIGO Document M1100296-v2}, 2011.

\bibitem{Abbott2016}
B.~P.~Abbot et~al.
\newblock {\em Living Reviews in Relativity}, 19, 2016.

\bibitem{Raffai2013}
P.~Raffai; L. Gond\'an; I. S. Heng; N. Kelecs\'enyi; J. Logue; Z. M\'arka;~S.
  M\'arka.
\newblock {\em Classical and Quantum Gravity}, 30:155004, 2013.

\bibitem{Schutz2011}
B.~F. Schutz.
\newblock {\em Classical and Quantum Gravity}, 28:125023, 2011.

\bibitem{WhitePaper2014}
The LVC.
\newblock {The LSC-Virgo White Paper on Gravitational Wave Searches and
  Astrophysics}.
\newblock {\em LIGO-T1400054, VIR-0176A-14}, 2014-2015 edition.

\bibitem{Aasi2014}
J.~Aasi et~al.
\newblock {\em The Astrophysical Journal}, 785:119, 2014.

\bibitem{Ming2016}
J.~Ming; B. Krishnan; M. A. Papa; C. Aulbert;~H. Fehrmann.
\newblock {\em Phys. Rev. D}, 93:064011, 2016.

\bibitem{Aasi2013b}
J.~Aasi et~al.
\newblock {\em Phys. Rev. D}, 88:102002, 2013.

\bibitem{Hanna2014}
C.~Hanna; I. Mandel;~W. Vousden.
\newblock {\em The Astrophysical Journal}, 784:8, 2014.

\bibitem{LCGT2009}
LCGT Special~Working Group.
\newblock {Study report on LCGT interferometer observation band}.
\newblock {\em JGW-T1000065-v1}, 2009.

\bibitem{InstWhitePaper2015}
The LSC.
\newblock {Instrument Science White Paper}.
\newblock {\em LIGO–T1400316}, 2015.

\bibitem{WhitePaper2013KAGRA}
Y.~Itoh; N. Kanda;~H. Tagoshi.
\newblock {KAGRA Data Analysis White Paper 2013}.
\newblock {\em JGW-T1402173-v1}, 2014.

\bibitem{Abdo2009}
A.~A.~Abdo et~al.
\newblock {\em The Astrophysical Journal Letters}, 699:L102, 2009.

\bibitem{Halpern2012}
J.~P. Halpern; E. V. Gotthelf;~F. Camilo.
\newblock {\em The Astrophysical Journal Letters}, 753:L14, 2012.

\bibitem{Kanner2012}
J.~Kanner; J. Camp; J. Racusin; N. Gehrels;~D. White.
\newblock {\em The Astrophysical Journal}, 759:22, 2012.

\bibitem{Bartos2012}
I.~Bartos; P. Brady;~S. M\'arka.
\newblock {\em Classical and Quantum Gravity}, 30:123001, 2013.

\bibitem{Metzger2011}
B.~D. Metzger and E.~Berger.
\newblock {\em The Astrophysical Journal}, 746:48, 2012.

\bibitem{Abadie2010}
J.~Abadie et~al.
\newblock {\em Classical and Quantum Gravity}, 27:173001, 2010.

\bibitem{Phinney1991}
E.~S. Phinney.
\newblock {\em The Astrophysical Journal}, 380:L17, 1991.

\bibitem{Postnov2014}
K.~A. Postnov and L.~R. Yungelson.
\newblock {\em Living Reviews in Relativity}, 17:3, 2014.

\bibitem{Kalogera2007}
V.~Kalogera; K. Belczynski; C. Kim; R. O'Shaughnessy;~B. Williams.
\newblock {\em Phys. Rept.}, 442:75, 2007.

\bibitem{Fong2010}
W.~Fong; E. Berger; D.~B. Fox.
\newblock {\em The Astrophysical Journal}, 708:9, 2010.

\bibitem{Bartos2015}
I.~Bartos; A. P. S. Crotts;~S. M\'arka.
\newblock {\em The Astrophysical Journal Letters}, 801:4, 2015.

\bibitem{Abadie2012}
LIGO~Scientific Collaboration and Virgo Collaboration.
\newblock {\em Astronomy and Astrophysics}, 539:15, 2012.

\bibitem{Igor2013}
I.~D. Karachentsev; D. I. Makarov; E.~I. Kaisina.
\newblock {\em The Astronomical Journal}, 145:101, 2013.

\bibitem{Freedman2001}
W.~L.~Freedman et~al.
\newblock {\em The Astrophysical Journal}, 553:47, 2001.

\bibitem{Mateo1998}
M.~Mateo.
\newblock {\em Annu. Rev. Astron. Astrophys.}, 36:435, 1998.

\bibitem{Patruel2003}
G.~Paturel; C. Petit; P. Prugniel; G. Theureau; J. Rousseau; M. Brouty; P.
  Dubois;~L. Cambr\'esy.
\newblock {\em Astronomy and Astrophysics}, 412:45, 2003.

\bibitem{Kopparapu2008}
R.~K. Kopparapu; C. Hanna; V. Kalogera; R. O'Shaughnessy; G. Gonz\'alez; P. R.
  Brady;~S. Fairhurst.
\newblock {\em The Astrophysical Journal}, 675:1459, 2008.

\bibitem{White2011}
D.~J. White; E. J. Daw; V.~S. Dhillon.
\newblock {\em Classical and Quantum Gravity}, 28:085016, 2011.

\bibitem{Sopuerta2015}
C.~F. Sopuerta.
\newblock {Gravitational Wave Astrophysics}.
\newblock {\em Springer International Publishing}, page~37, 2015.

\bibitem{Belczynski2007}
K.~Belczynski; R. E. Taam; V. Kalogera; F. A. Rasio;~T. Bulik.
\newblock {\em The Astrophysical Journal}, 662:504, 2007.

\bibitem{Cutler1994}
C.~Cutler and {\'E}.~E. Flanagan.
\newblock {\em Phys. Rev. D.}, 49:2658, 1994.

\end{thebibliography}

\end{document}